\documentstyle[12pt,epsf,rotate]{article}

\def\Lms{\Lambda_{\overline{MS}}} 
 
\def\R1{\varepsilon_1}
\def\E8{\varepsilon_8}

\def\as{\alpha_s}

\newcommand{\mb}{m_{\rm b}}

\newcommand{\mz}{M_{\rm Z}}
\newcommand{\gev}{\, {\rm GeV}}
\newcommand{\mev}{\, {\rm MeV}}
\newcommand{\bea}{\begin{eqnarray}}
\newcommand{\eea}{\end{eqnarray}}
\newcommand{\be}{\begin{equation}}
\newcommand{\ee}{\end{equation}}
\newcommand{\bi}{\begin{itemize}}
\newcommand{\ei}{\end{itemize}}
\newcommand{\ord}{{\cal O}}

\begin{document}
\thispagestyle{empty}
\begin{flushright}
TUM-HEP-315/98 \\
hep-ph/9806278 \\
 June 1998
\end{flushright}
\vskip1truecm
\centerline{\Large\bf  Generalized Factorization in Non-leptonic}
\centerline{\Large\bf Two-Body B-Decays: a Critical Look
   \footnote[1]{\noindent
   Supported by the German
   Bundesministerium f\"ur Bildung und Forschung under contract
   06 TM 874 and by the DFG project Li 519/2-2.}}
\vskip1truecm
\centerline{\large\bf Andrzej J. Buras and Luca Silvestrini}
\bigskip
\centerline{\sl Technische Universit\"at M\"unchen, Physik Department}
\centerline{\sl D-85748 Garching, Germany}
\vskip1truecm
\centerline{\bf Abstract}
We reanalyze critically the generalized factorization hypothesis
in non-leptonic two-body B-decays discussed recently by several authors.
In particular we address the determination of the factorization
scale $\mu_f$ and of the non-perturbative parameters
$\xi^{\rm NF}_1(\mb)$ and $\xi^{\rm NF}_2(\mb)$ which are supposed to
measure non-factorizable contributions to hadronic matrix elements
with $\xi^{\rm NF}_i(\mu_f)=0.$ We emphasize that both $\mu_f$ and
$\xi^{\rm NF}_i(\mb)$ are renormalization scheme dependent and
we demonstrate analytically and numerically that for any chosen
scale $\mu_f=\ord(\mb)$ it is possible to find a renormalization
scheme for which $\xi^{\rm NF}_1(\mu_f)=\xi^{\rm NF}_2(\mu_f)=0.$
The existing data indicate that such ``factorization schemes"
differ from the commonly used schemes NDR and HV. Similarly
we point out that the recent extractions of the effective number
of colours $N^{\rm eff}$ from two-body non-leptonic B-decays
while $\mu$ and renormalization scheme independent suffer from
gauge dependences and infrared dependences.

\vfill
\newpage
 
\section{Introduction}
Two-body non-leptonic B-decays play an important role in the
phenomenology of weak decays not only probing the structure of weak
interactions corrected by short distance QCD effects but also
providing some insight into the non-perturbative phenomena related to
long distances. The increasing experimental information on these
decays, in particular from the CLEO detector, stimulated recently
several new theoretical analyses of these decays.  The most extensive
analyses of this type are based on the factorization of hadronic
matrix elements of local operators \cite{FEYNMAN}--\cite{NEUBERT}
which has recently been extended to the so-called generalized
factorization hypothesis \cite{Cheng}--\cite{AKL98}.

In the strict factorization approach two-body decays are parametrized
in terms of two phenomenological parameters $a_1$ and $a_2$ \cite{BAUER}
which in
QCD are given by
\begin{equation}\label{BS23}
a_1(\mu)=C_1(\mu)+\frac{1}{N} C_2(\mu)~, \qquad
a_2(\mu)=C_2(\mu)+\frac{1}{N} C_1(\mu).
\end{equation}
Here $C_{1,2}(\mu)$ are the short distance Wilson coefficient functions
of the relevant current-current operators $O_{1,2}$ for which explicit
expressions will be given below. $N$ is the number of colours with
$N=3$ in QCD.

One
distinguishes then three classes of decays for which the amplitudes have the
following general structure \cite{BAUER,NEUBERT}:
\begin{equation}\label{1}
A_{\rm I}=\frac{G_F}{\sqrt{2}} V_{CKM}a_1(\mu)\langle O_1\rangle_F 
\qquad {\rm (Class~I)}
\end{equation}
\begin{equation}\label{2}
A_{\rm II}=\frac{G_F}{\sqrt{2}} V_{CKM}a_2(\mu)\langle O_2\rangle_F 
\qquad {\rm (Class~II)}
\end{equation}
\begin{equation}\label{3N}
A_{\rm III}=
\frac{G_F}{\sqrt{2}} V_{CKM}[a_1(\mu)+x a_2(\mu)]\langle O_1\rangle_F
 \qquad {\rm (Class~III)}
\end{equation}
Here $V_{CKM}$ denotes symbolically the CKM factor characteristic for a
given decay. 
$\langle O_i\rangle_F$ are factorized 
hadronic matrix
elements of the operators $O_i$ given as products of matrix elements of
quark currents and $x$ is a non-perturbative factor equal to unity in
the flavour symmetry limit. 

The simplicity of this approach is very appealing.
Once the matrix elements $\langle O_i\rangle_F $
have been expressed in terms of various meson
decay constants and generally model dependent form factors, predictions
for non-leptonic heavy meson decays can be made. 
An incomplete list of analyses of this type is given in 
\cite{BAUER,NEUBERT,LNF} and will be extended below.
 
On the other hand,
it is well known that
non-factorizable contributions must be present in the hadronic matrix
elements of the current-current operators $O_1$ and $O_2$ in order
to cancel the $\mu$ dependence of $C_i(\mu)$ or $a_i(\mu)$ so that
the physical amplitudes do not depend on the arbitrary renormalization
scale $\mu$. 
$\langle O_i\rangle_F$ being products of matrix elements of
conserved currents
are $\mu$-independent and the cancellation of the $\mu$ dependence
in (\ref{1})--(\ref{3N}) does not take place.
Consequently from the point of view of QCD
the factorization approach can be at best correct at a single value
of $\mu$, the so-called factorization scale $\mu_f$. Although the
approach itself does not provide the value of $\mu_f$, the  
proponents
of factorization expect $\mu_f=O(m_b)$ and $\mu_f=O(m_c)$ for
B-decays and D-decays respectively.

The fact that $\langle O_i\rangle_F$ are $\mu$-independent but
$a_i(\mu)$ are $\mu$-dependent, which is clearly inconsistent,
inspired a number of authors \cite{Cheng}--\cite{AKL98} to
generalize the concept of factorization. 

In the formulation
due to Neubert and Stech \cite{NS97} the $\mu$-dependent parameters
$a_1(\mu)$ and $a_2(\mu)$ are replaced by $\mu$-independent
effective parameters $a_1^{\rm eff}$ and $a_2^{\rm eff}$.
The latter depend on $C_i(\mu)$ and two non-perturbative parameters
$\varepsilon_1(\mu)$ and $\varepsilon_8(\mu)$ which parametrize
the non-factorizable contributions to
the hadronic matrix elements of the operators $O_{1,2}$. 
In the case of strict
factorization $\varepsilon_i$ vanish and $a_{1,2}^{\rm eff}$
reduce to $a_{1,2}(\mu)$. The $\mu$ dependence of $\varepsilon_i(\mu)$
cancels the $\mu$-dependence of $C_{1,2}(\mu)$ so that 
$a_{1,2}^{\rm eff}$ are indeed scale independent.

From the phenomenological point of view there is no change here
relative to the standard factorization as only $a_i(\mu)$ have
been replaced by $a^{\rm eff}_i$ in the formulae
(\ref{1})--(\ref{3N}). On the other hand, as stressed 
in \cite{NS97}, the new formulation should allow in 
principle some insight into the importance of non-factorizable
contributions to hadronic matrix elements.

In this context we should remark that in the recent literature mainly
the $\mu$-dependence of the non-factorizable contributions has been
emphasized. Their renormalization scheme dependence has often not been
discussed. It is the latter issue which will be important in the
discussion below.  Indeed at the next-to-leading level in the
renormalization group improved perturbation theory the coefficients
$C_i(\mu)$ depend on the renormalization scheme for operators. Again
only the presence of non-factorizable scheme dependent contributions
in $\langle O_i\rangle$ can remove this scheme dependence in the
physical amplitudes and in particular in $a^{\rm eff}_i$.  The
renormalization scheme dependence emphasized here, and discussed in
the context of strict factorization in \cite{AJBNF}, is rather
annoying from the factorization point of view as it precludes a unique
phenomenological determination of $\mu_f$ as we will show explicitly
below. In particular we will demonstrate that for any chosen scale
$\mu_f=\ord(\mb)$ it is possible to find a renormalization scheme for
which the non-factorizable parameters $\varepsilon_{1,8}(\mu_f)$
simultaneously vanish. This finding casts some doubts on the
usefulness of the formulation in \cite{NS97} with respect to the study of
non-factorizable contributions to non-leptonic decays.

The generalized factorization  presented in \cite{Cheng,GNF,AKL98} is
similar in spirit but includes more dynamics than the formulation
in \cite{NS97}. Here the non-factorizable contributions to the
matrix elements are calculated in a perturbative framework at the
one-loop level. Subsequently these non-factorizable contributions
are combined with the coefficients $C_i(\mu)$ to obtain effective
$\mu$ and renormalization scheme independent coefficients 
$C_i^{\rm eff}$. The effective parameters $a^{\rm eff}_i$
are given in this formulation as follows:
\begin{equation}\label{BS23F}
a_1^{\rm eff}=C^{\rm eff}_1+\frac{1}{N^{\rm eff}} C^{\rm eff}_2 \qquad
a_2^{\rm eff}=C^{\rm eff}_2+\frac{1}{N^{\rm eff}} C^{\rm eff}_1
\end{equation}
with analogous expressions for $a_{i}^{\rm eff}$ ($i=3-10$) parametrizing
penguin contributions.
Here $N^{\rm eff}$ is treated as a phenomenological parameter which
models the non-factorizable contributions to the hadronic matrix elements.
In particular it has been suggested in \cite{Cheng,GNF,AKL98} that
the values for $N^{\rm eff}$ extracted from the data on two-body
non-leptonic decays should teach us about the pattern of 
non-factorizable contributions.

Unfortunately, as we will demonstrate below, also this approach has
its weak points. Although $C^{\rm eff}_{1,2}$ are $\mu$ and
renormalization scheme independent, they are both gauge and infrared
regulator dependent. The latter dependences originate in the
perturbative evaluation of the scheme dependent finite contributions
to the matrix elements, needed for the cancellation of the
renormalization scheme dependence of $C_i(\mu)$.  Consequently,
whereas the extracted $N^{\rm eff}$ is renormalization scheme and
renormalization scale independent, it is a gauge and infrared
regulator dependent quantity. This finding casts some doubts on the
usefulness of the formulation in \cite{Cheng,GNF,AKL98} with respect
to the study of non-factorizable contributions to non-leptonic decays.

The rest of our paper amounts to putting all these statements in explicit
terms. In section 2 we review the approach in \cite{NS97}.
In section 3 we reformulate this approach by replacing the parameters
$\varepsilon_{1,8}(\mu)$ by two new parameters $\xi^{\rm NF}_{1,2}$.
This reformulation allows us to make our points with regard to \cite{NS97}
in a more transparent manner. In particular we derive general expressions
which allow to find, for a given $\mu_f$, the renormalization scheme
in which $\varepsilon_{1,8}(\mu)$ or $\xi^{\rm NF}_{1,2}(\mu)$ 
simultaneously
vanish. In section 4 we illustrate our points with a few numerical
examples and in section 5 we make a critical analysis of the approach
in \cite{Cheng,GNF,AKL98}. We end our paper with a brief summary and 
conclusions.

\section{Generalized Factorization}
In order to describe generalized factorization in explicit terms
let us consider the decay
$\bar B^0\to D^+\pi^-$. Then the
relevant effective Hamiltonian is given by
\begin{equation}\label{BS1}
H_{\rm eff}=\frac{G_F}{\sqrt{2}}V_{cb}V_{ud}^{*}
\lbrack C_1(\mu) O_1+C_2(\mu)O_2 \rbrack~,
\end{equation}
where
\begin{equation}\label{BS2}
O_1=(\bar d_\alpha u_\alpha)_{V-A} (\bar c_\beta b_\beta)_{V-A}~,
\qquad 
O_2=(\bar d_\alpha u_\beta)_{V-A} (\bar c_\beta b_\alpha)_{V-A}
\end{equation}
with $(\alpha,\beta=1,2,3)$ denoting colour indices and $V-A$ referring to
$\gamma_\mu (1-\gamma_5)$. $C_1(\mu)$ and $C_2(\mu)$ are short distance
Wilson coefficients computed at the renormalization scale $\mu=O(m_b)$.
Note that we use here the labelling of the operators as given in
\cite{BAUER,NEUBERT} which differs from 
\cite{BJL}--\cite{ROME} by the interchange
$1\leftrightarrow 2$.
Since all four quark flavours entering the operators in (\ref{BS2})
are different from each other, no penguin operators contribute to
this decay. 

Using Fierz reordering and colour identities one can rewrite the
amplitude for $\bar B^0\to D^+\pi^-$ as
\begin{equation}\label{BS3}
A(\bar B^0\to D^+\pi^-)=\frac{G_F}{\sqrt{2}}V_{cb}V_{ud}^{*}
a^{\rm eff}_1 \langle O_1\rangle_F~,
\end{equation}
where
\be
\langle O_1\rangle_F=
\langle\pi^-\mid(\bar d u)_{V-A}\mid 0\rangle
\langle D^+\mid (\bar c b)_{V-A}\mid \bar B^0\rangle
\ee
is the factorized matrix element of the operator $O_1$ and
summation over colour indices in each current is understood.

The effective parameter $a^{\rm eff}_1$ is given by \cite{NS97}
\be\label{BS4}
a^{\rm eff}_1=\left(C_1(\mu)+\frac{1}{N} C_2(\mu)\right)
[1+\varepsilon_1^{(BD,\pi)}(\mu)]+C_2(\mu)\varepsilon_8^{(BD,\pi)}(\mu).
\ee
$\varepsilon_1^{(BD,\pi)}(\mu)$ and $\varepsilon_8^{(BD,\pi)}(\mu)$
are two hadronic parameters defined by \cite{NS97}
\be\label{BS5}
\varepsilon_1^{(BD,\pi)}(\mu)\equiv
\frac{\langle \pi^-D^+|(\bar d u)_{V-A}(\bar c b)_{V-A}|\bar B^0\rangle}
{\langle O_1 \rangle_F}-1
\ee
and
\be\label{BS6}
\varepsilon_8^{(BD,\pi)}(\mu)\equiv 2
\frac{\langle \pi^-D^+|(\bar dt_a u)_{V-A}
(\bar ct_a b)_{V-A}|\bar B^0\rangle}
{\langle O_1\rangle_F}
\ee
with $t_a$ denoting the colour matrices in the standard Feynman
rules. $\varepsilon_i(\mu)$ parametrize the non-factorizable 
contributions to
the hadronic matrix elements of operators. In the case of strict
factorization $\varepsilon_i$ vanish.

It should be emphasized that no approximation has been made
in (\ref{BS3}). Since the matrix element $\langle O_1 \rangle_F$
is scale and renormalization scheme independent this must also
be the case for the effective coefficient $a^{\rm eff}_1$.
Indeed the scale and scheme dependences of the coefficients
$C_1(\mu)$ and $C_2(\mu)$ are cancelled by those present in
the hadronic parameters $\varepsilon_i(\mu)$. We will give
explicit formulae for these dependences below.

A similar exercise with the
amplitude for $\bar B^0\to D^0\pi^0$ gives
\begin{equation}\label{BS7}
A(\bar B^0\to D^0\pi^0)=\frac{G_F}{\sqrt{2}}V_{cb}V_{ud}^{*}
a^{\rm eff}_2 \langle O_2 \rangle_F~,
\end{equation}
where
\be\label{fact2}
\langle O_2 \rangle_F=
\langle D^0\mid(\bar c u)_{V-A}\mid 0\rangle
\langle \pi^0\mid (\bar d b)_{V-A}\mid \bar B^0\rangle
\ee
is the factorized matrix element of the operator $O_2$.

The effective parameter $a^{\rm eff}_2$ is given by \cite{NS97}
\be\label{BS8}
a^{\rm eff}_2=\left(C_2(\mu)+\frac{1}{N} C_1(\mu)\right)
[1+\varepsilon_1^{(B\pi,D)}(\mu)]+C_1(\mu)\varepsilon_8^{(B\pi,D)}(\mu)~.
\ee
$\varepsilon_1^{(B\pi,D)}(\mu)$ and $\varepsilon_8^{(B\pi,D)}(\mu)$
are two hadronic parameters defined by
\be\label{BS9}
\varepsilon_1^{(B\pi,D)}(\mu)\equiv
\frac{\langle \pi^0D^0|(\bar c u)_{V-A}(\bar d b)_{V-A}|\bar B^0\rangle}
{\langle O_2 \rangle_F}-1
\ee
and
\be\label{BS10}
\varepsilon_8^{(B\pi,D)}(\mu)\equiv 2
\frac{\langle \pi^0D^0|(\bar ct_a u)_{V-A}
(\bar d t_a b)_{V-A}|\bar B^0\rangle}
{\langle O_2 \rangle_F}
\ee
Again the $\mu$ and scheme dependences of $\varepsilon_i$ in
(\ref{BS9}) and (\ref{BS10}) cancel the corresponding
dependences in $C_i(\mu)$ so that the effective coefficient
$a^{\rm eff}_2$ is $\mu$ and scheme independent.

Following section 5.1 of \cite{BJL} it is straightforward to
find the explicit $\mu$ and scheme dependences of the hadronic
parameters $\varepsilon_i(\mu)$. To this end we note
that the $\mu$ dependence of the matrix elements of the
operators $O_\pm=(O_1\pm O_2)/2$ is given by \cite{BJL}
\be\label{BS110}
\langle O_\pm(\mu)\rangle = U_\pm(\mb,\mu) \langle O_\pm(\mb)\rangle
\ee
where the evolution function $U_\pm(\mb,\mu)$ 
including NLO QCD corrections is given
by
\begin{equation}\label{BS11}
U_\pm(\mb,\mu)=\left[1+\frac{\alpha_s(\mb)}{4\pi}J_\pm\right]
      \left[\frac{\alpha_s(\mu)}{\alpha_s(\mb)}\right]^{d_\pm}
\left[1-\frac{\alpha_s(\mu)}{4\pi}J_\pm\right]
\end{equation}
with
\begin{equation}\label{BS12}
J_\pm=\frac{d_\pm}{\beta_0}\beta_1-\frac{\gamma^{(1)}_\pm}{2\beta_0}~,
\qquad\qquad
d_\pm=\frac{\gamma^{(0)}_\pm}{2\beta_0}~,
\end{equation}
\begin{equation}\label{BS13}
\gamma^{(0)}_\pm=\pm 2 (3\mp 1)~,
\qquad\quad
\beta_0=11-\frac{2}{3}f~,
\qquad\quad
\beta_1=102-\frac{38}{3}f~,
\end{equation}
\begin{equation}\label{BS14}
\gamma^{(1)}_{\pm}=\frac{3 \mp 1}{6}
\left[-21\pm\frac{4}{3}f-2\beta_0\kappa_\pm\right].
\end{equation}
Here $\kappa_\pm$, introduced in \cite{AJBNF},  
distinguishes between various renormalization
schemes:
\begin{equation}\label{BS16}
\kappa_\pm = \left\{ \begin{array}{rcc}
0 & (\rm{NDR}) & \cite{WEISZ}  \\
\mp 4 & (\rm{HV}) & \cite{WEISZ} \\
\mp 6-3 & (\rm{DRED})& \cite{ALT}
\end{array}\right.
\end{equation}

Thus $J_\pm$ in (\ref{BS12}) can also be written as
\begin{equation}\label{BS17}
J_\pm=(J_\pm)_{NDR}+\frac{3\mp 1}{6}\kappa_\pm
=(J_\pm)_{NDR}\pm\frac{\gamma^{(0)}_\pm}{12}\kappa_\pm~.
\end{equation}
The 
$\overline{MS}$ coupling \cite{BBDM} is given by
\begin{equation}\label{BS18}
\alpha_s(\mu)=\frac{4\pi}{\beta_0 \ln(\mu^2/\Lambda^2_{\overline{MS}})}
\left[1-\frac{\beta_1}{\beta^2_0}
\frac{\ln\ln(\mu^2/\Lambda^2_{\overline{MS}})}
{\ln(\mu^2/\Lambda^2_{\overline{MS}})}\right]~.
\end{equation}
The formulae given above
 depend on $f$, the number of active flavours. In the case of
B-decays $f=5$. The present world average for $\alpha_s(M_Z)$ is 
\cite{WEBER}:
\begin{equation}\label{BS19}
\alpha_s(M_Z)=0.118\pm0.003
\qquad\quad
\Lambda_{\overline{MS}}^{(5)}=(225\pm 40)~MeV
\end{equation}
where the superscript stands for $f=5$.

Having these formulae at hand it is straightforward to show
that the $\mu$-dependence of $\varepsilon_1(\mu)$ and
$\varepsilon_8(\mu)$ is governed by the following equations:
\begin{eqnarray}\label{BS20}
1+\varepsilon_1(\mu)&=&
\frac{1}{2}
\left[\left(1+\frac{1}{N}\right)[1+\R1(\mb)]+\E8(\mb)\right]
U_+(\mb,\mu)\\
&+&
\frac{1}{2}
\left[\left(1-\frac{1}{N}\right)[1+\R1(\mb)]-\E8(\mb)\right]
U_-(\mb,\mu)~,   \nonumber
\end{eqnarray}

\begin{eqnarray}\label{BS21}
\varepsilon_8(\mu)&=&
\frac{1}{2}
\left[\left(1-\frac{1}{N}\right)\E8(\mb)+
\left(1-\frac{1}{N^2}\right)[1+\R1(\mb)]\right]
U_+(\mb,\mu)\\
&+&
\frac{1}{2}
\left[\left(1+\frac{1}{N}\right)\E8(\mb)-
\left(1-\frac{1}{N^2}\right)[1+\R1(\mb)]\right]
U_-(\mb,\mu)~.
 \nonumber
\end{eqnarray}

These formulae reduce to the ones given in \cite{NS97} when $J_\pm$ in
(\ref{BS11}) are set to zero. They give both the $\mu$-dependence
and renormalization scheme dependence of $\varepsilon_i$. The
latter dependence has not been considered in \cite{NS97}.

\section{A Different Formulation}
In order to be able to discuss the relation of our work to the one of
\cite{NS97} we have used until now, as in \cite{NS97}, the hadronic
parameters $\R1(\mu)$ and $\E8(\mu)$ to describe non-factorizable
contributions. On the other hand, it appears
to us that it is more convenient to work instead with two other
parameters defined simply by
\be\label{BS22}
a^{\rm eff}_1=a_1(\mu)+\xi^{\rm NF}_1(\mu),
\quad\quad
a^{\rm eff}_2=a_2(\mu)+\xi^{\rm NF}_2(\mu)~,
\ee
where $a_1(\mu)$ and $a_2(\mu)$ are given in (\ref{BS23}).

Comparison with (\ref{BS4}) and (\ref{BS8}) gives
\be\label{xi1}
\xi^{\rm NF}_1(\mu)=\R1(\mu) a_1(\mu)+\E8(\mu) C_2(\mu)~,
\ee
\be\label{xi2}
\xi^{\rm NF}_2(\mu)=\bar\R1(\mu) a_2(\mu)+\bar\E8(\mu) C_1(\mu)~,
\ee
where
\be\label{E18}
\R1(\mu)=\R1^{(BD,\pi)}, \qquad \E8(\mu)=\E8^{(BD,\pi)}~,
\ee
\be\label{E19}
\bar\R1(\mu)=\R1^{(B\pi,D)}, \qquad \bar\E8(\mu)=\E8^{(B\pi,D)}~.
\ee
and
$a_i(\mu)$, given in (\ref{BS23}), are the parameters used in the 
framework of the
strict factorization hypothesis in which $\xi^{\rm NF}_i(\mu)$
are set to zero. Their $\mu$ and renormalization scheme dependence
can be studied using
\begin{equation}\label{10}
C_1(\mu)=\frac{z_+(\mu)+z_-(\mu)}{2}~,
\qquad\qquad
C_2(\mu)=\frac{z_+(\mu)-z_-(\mu)}{2}~,
\end{equation}
where
\begin{equation}\label{11}
z_\pm(\mu)=\left[1+\frac{\alpha_s(\mu)}{4\pi}J_\pm\right]
      \left[\frac{\alpha_s(M_W)}{\alpha_s(\mu)}\right]^{d_\pm}
\left[1+\frac{\alpha_s(M_W)}{4\pi}(B_\pm-J_\pm)\right]
\end{equation}
with
\begin{equation}\label{15}
B_\pm=\frac{3 \mp 1}{6}\left[\pm 11+\kappa_\pm\right]
\end{equation}
and all other quantities defined before. The $\mu$ and scheme
dependences of $\xi_i^{\rm NF}$ can in principle be found by using the
dependences of $C_i(\mu)$ given above and $\varepsilon_i(\mu)$ in
(\ref{BS20}) and (\ref{BS21}).  To this end, however, one needs the
determination of the non-perturbative parameters $\varepsilon_i(\mu)$
and $\bar\varepsilon_i(\mu)$ at a single value of $\mu$. If, as done
in \cite{NS97}, $a_i^{\rm eff}$ are assumed to be universal
parameters, the determination of $\varepsilon_i(\mu)$ and
$\bar\varepsilon_i(\mu)$ is only possible if one also makes the
following {\it universality} assumptions:
\be\label{FACTU}
\R1(\mu)=\bar\R1(\mu), \qquad \E8(\mu)=\bar\E8(\mu)~.
\ee
In \cite{NS97} such an assumption was not necessary since
$\varepsilon_1(\mu)$, $\bar\varepsilon_1(\mu)$ and
$\varepsilon_8(\mu)$ were neglected, and only $\bar\varepsilon_8(\mu)$
was kept in the analysis. 

With the assumptions in (\ref{FACTU}), $\R1(\mu)$ and $\E8(\mu)$
can indeed be found once the effective parameters $a_i^{\rm eff}$
have been determined experimentally. Using (\ref{BS4}) and
(\ref{BS8}) together with (\ref{FACTU}) we find
\be\label{E1MU}
\R1(\mu)=\frac{C_1(\mu) a_1^{\rm eff}-C_2(\mu)a_2^{\rm eff}}
          {C^2_1(\mu)-C^2_2(\mu)}-1
\ee
\be\label{E2MU}
\E8(\mu)=\frac{a_2^{\rm eff}}{C_1(\mu)}-
\left(\frac{C_2(\mu)}{C_1(\mu)}+\frac{1}{N}\right) [1+\R1(\mu)]
\ee
On the other hand  $\xi_i^{\rm NF}(\mu)$ can be determined
without the universality assumption (\ref{FACTU}) from two decays 
simply as follows
\be\label{BS24}
 \xi^{\rm NF}_1(\mu)=a^{\rm eff}_1-a_1(\mu)
\quad\quad
 \xi^{\rm NF}_2(\mu)=a^{\rm eff}_2-a_2(\mu)
\ee
We will analyze the formulae (\ref{E1MU}), (\ref{E2MU}) and (\ref{BS24})
in the next section.

The formulae in (\ref{BS24}) make it clear that the strict factorization
in which $\xi_i^{\rm NF}(\mu)$ vanish can be at best correct at a single
value of $\mu$, the so-called factorization scale $\mu_f$. In the
first studies of factorization $\mu_f=\mb$ 
has been assumed. It has been concluded that such a choice is
not in accord with the data \cite{BAUER,LNF,NEUBERT}.

The idea of the generalized factorization as formulated by Neubert and
Stech \cite{NS97} (see also \cite{Cheng,Soares} for earlier presentations)
is to allow $\mu_f$ to be different from $\mb$ and to extract
first the non-factorizable parameters $\varepsilon_i(\mb)$ from the
data. Subsequently the factorization scale $\mu_f$ can be found
by requiring these parameters to vanish.

In the numerical analysis of this procedure done in \cite{NS97}
a further assumption has been made. Using large $N$ arguments
it has been argued that $\R1(\mu)$ can be set to zero while
$\E8(\mu)$ can be sizable. The resulting expressions
for $a_i^{\rm eff}$ are then
\be\label{NSF}
a_1^{\rm eff}=C_1(\mb), 
\qquad a_2^{\rm eff}=a_2(\mb)+C_1(\mb)\E8(\mb)
\ee
where additional small terms have been dropped in order to obtain
the formula for $a_1^{\rm eff}$. Using subsequently
the extracted value $a_2^{\rm eff}=0.21\pm0.05$ together with
the  coefficients $C_i(\mb)$ from \cite{WEISZ} one finds
$\E8(\mb)=0.12\pm0.05$ \cite{NS97}. Next assuming $\E8(\mu_f)=0$ 
one can find the factorization scale $\mu_f$
by inverting the formula
\be\label{UF}
\E8(\mb)=-\frac{4\as(\mb)}{3\pi}\ln\frac{\mb}{\mu_f}
\ee
which follows from (\ref{BS21}) with $\E8(\mu_f)=0$ and $\R1(\mb)=0.$
Thus
\be\label{UF1}
 \mu_f=\mb \exp\left[\frac{3\pi\E8(\mb)}{4\as(\mb)}\right].
\ee
Taking $\mb=4.8~\gev$ and $\as(m_b)=0.21$ 
(corresponding to $\as(\mz)=0.118$) we find using $\E8(\mb)=0.12\pm0.05$
a rather large factorization scale $\mu_f=(15.9^{+11.3}_{-6.6})~\gev$,
roughly a factor of 3-4 higher than $\mb$. This implies that
non-factorizable contributions in hadronic matrix elements at scales
close to $\mb$ are sizable. This is also signaled by the 
value of $\E8(\mb)\approx0.12$ which is as large as the factorizable
contribution $a_2(\mb)=0.10$ to the effective parameter
$a_2^{\rm eff}=0.21\pm0.05$.

We would like to emphasize that such an interpretation of the
analysis of Neubert and Stech \cite{NS97} would be
misleading. As stressed in \cite{AJBNF} the coefficient $a_2(\mu)$
is very strongly dependent on the renormalization scheme.
Consequently for a given value of $a_2^{\rm eff}$ also
$\xi_2^{NF}(\mb)$ and $\E8(\mb)$ are strongly scheme dependent.
This shows \cite{AJBNF}, that a meaningful analysis of the
$\mu$-dependences in non-leptonic decays, such as the search for the
factorization scale $\mu_f$, cannot be be made without simultaneously
considering the scheme dependence. This is evident if one recalls that
any variation of $\mu_f$ in the leading logarithm is equivalent to
a shift in constant non-logarithmic terms. The latter represent
NLO contributions in the renormalization group improved
perturbation theory and must be included for a meaningful extraction
of $\mu_f$ or any other scale like $\Lms$. However, once  the NLO
contributions are taken into account, the renormalization scheme
dependence enters the analysis and consequently the factorization
scale $\mu_f$ at which the non-factorizable hadronic parameters
$\xi_i^{NF}(\mu_f)$ or $\varepsilon_i(\mu_f)$ vanish is renormalization
scheme dependent.  
  
From this discussion it becomes clear that for any chosen scale
$\mu_f=\ord(\mb)$, it is always possible to find a renormalization
scheme for which
\be\label{xifac}
\xi_1^{NF}(\mu_f)=\xi_2^{NF}(\mu_f)=0~.
\ee
Indeed as seen in (\ref{BS24}) $\xi_i^{NF}(\mu)$ depend through
$a_i(\mu)$ on $\kappa_\pm$
which characterize a given renormalization scheme. The choice
of $\kappa_\pm$
corresponds to a particular finite renormalization of the operators
$O_\pm$ in addition to the renormalization in the NDR scheme. It
is then straightforward to find the values of $\kappa_\pm$ which
assure that for a chosen scale $\mu_f$ the conditions in (\ref{xifac})
are satisfied. We find
\be\label{kappa+}
\kappa_+=3
\left[\frac{3}{4}\frac{a_1^{\rm eff}+a_2^{\rm eff}}{W_+(\mu_f)}-1\right] 
\frac{4\pi}{\as(\mu_f)}-3 (J_+)_{\rm NDR}~,
\ee
\be\label{kappa-}
\kappa_-=\frac{3}{2}
\left[\frac{3}{2}\frac{a_1^{\rm eff}-a_2^{\rm eff}}{W_-(\mu_f)}-1\right] 
\frac{4\pi}{\as(\mu_f)}-\frac{3}{2} (J_-)_{\rm NDR}~,
\ee
where
\be\label{W+-}
W_\pm(\mu_f)=\left[\frac{\alpha_s(M_W)}{\alpha_s(\mu_f)}\right]^{d_\pm}
\left[1+\frac{\alpha_s(M_W)}{4\pi}(B_\pm-J_\pm)\right]
\ee
with $(J_\pm)_{\rm NDR}$ being the values of $J_\pm$ in the NDR scheme.
$W_\pm(\mu_f)$ are clearly renormalization scheme independent as
$B_\pm-J_\pm$ are scheme independent.
\section{Numerical Analysis}
Before presenting the numerical analysis of the formulae derived in
the preceding section, we would like to clarify the difference
between the Wilson coefficients in (\ref{10}) and (\ref{11}) used by
us and the ones employed in \cite{NS97}. In \cite{NS97} the scheme
independent coefficients $\tilde z_\pm(\mu)$ of \cite{WEISZ}  
instead of $z_\pm(\mu)$ have been used. These are obtained by
multiplying $z_\pm(\mu)$ by $(1-B_\pm \alpha_s(\mu)/4\pi)$ 
so that
\begin{equation}\label{T11}
\tilde z_\pm(\mu)=
      \left[\frac{\alpha_s(M_W)}{\alpha_s(\mu)}\right]^{d_\pm}
\left[1+\frac{\alpha_s(M_W)-\alpha_s(\mu)}{4\pi}(B_\pm-J_\pm)\right].
\end{equation}
These coefficients are clearly not the coefficients of the operators
$O_\pm$ used in \cite{NS97} and here. In order to be consistent the
matrix elements $\langle O_\pm \rangle$ should then be replaced by
\be\label{TOPM}
\langle \tilde O_\pm \rangle=
(1+B_\pm \alpha_s(\mu)/4\pi) \langle O_\pm \rangle.
\ee
This explains why the results of our numerical analysis differ
considerably from the ones presented in \cite{NS97}. 
We strongly advice the practitioners of
non-leptonic decays not to use the scheme independent coefficients
of \cite{WEISZ} in phenomenological applications. These coefficients
have been introduced to test the compatibility of different 
renormalization schemes and can only be used for phenomenology
together with $\langle \tilde O_\pm \rangle$. This would however
unnecessarily complicate the analysis and it is therefore advisable
to  work with the true coefficients $C_i(\mu)$ of the operators $O_i$ 
as given   in (\ref{10}) and (\ref{11}).

\begin{figure}   
    \begin{center}
\input{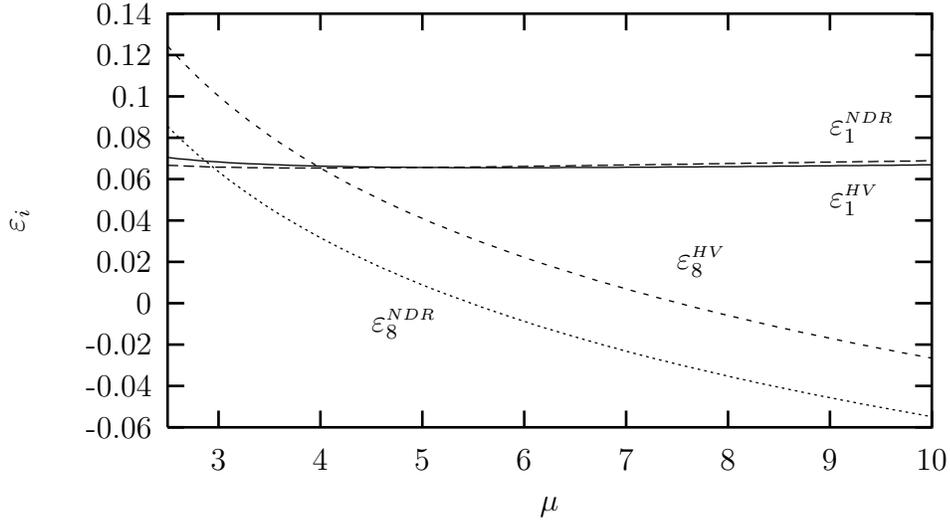}
    \end{center}
    \caption[]{$\varepsilon_{1,8}(\mu)$ in the NDR and HV schemes.}
    \label{SILV1}
\end{figure}

\begin{figure}   
    \begin{center}
\input{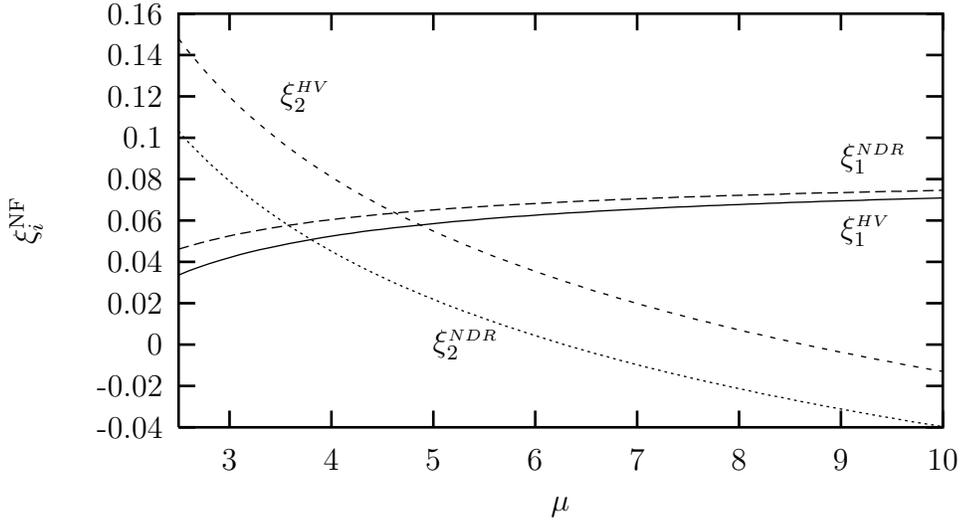}
    \end{center}
    \caption[]{$\xi^{\rm NF}_{1,2}(\mu)$ in the NDR and HV schemes.}
    \label{SILV2}
\end{figure}

In \cite{NS97} the following values of $a_i^{\rm eff}$ have been
extracted from existing data on two-body B-decays
\be\label{aiexp}
a_1^{\rm eff}=1.08\pm0.04 \qquad a_2^{\rm eff}=0.21\pm 0.05
\ee
with similar results given in
\cite{Cheng,Soares,LNF,GNF,AKL98}.
In order to illustrate various points made in the preceding section,
we take the central values of $a_i^{\rm eff}$ in (\ref{aiexp}).
Using (\ref{E1MU})-(\ref{BS24}) we calculate $\varepsilon_i(\mu)$
and $\xi_i^{NF}(\mu)$ as a function of $\mu$ in the range
$2.5~\gev \le \mu \le 10~\gev$ for the NDR and HV schemes. The
results are shown in fig.~\ref{SILV1} and fig.~\ref{SILV2}.
We observe that $\varepsilon_1(\mu)$ and 
$\xi^{\rm NF}_1(\mu)$ are only weakly $\mu$ and scheme
dependent in accordance with the findings in \cite{AJBNF},
where these dependences have been studied for $a_i(\mu)$
defined in (\ref{BS23}). The strong $\mu$ and scheme dependences
of $a_2(\mu)$ found there translate into similar strong dependences
of $\varepsilon_8(\mu)$ and $\xi_2^{\rm NF}(\mu)$.

We make the following observations:
\bi
\item
$\varepsilon_1(\mu)$ and $\xi_1^{\rm NF}(\mu)$ are non-zero in the
full range of $\mu$ considered.
\item
$\varepsilon_8(\mu)$ and $\xi_2^{\rm NF}(\mu)$ vary strongly with
$\mu$ and vanish in the NDR scheme for $\mu=5.5~\gev$ and 
$\mu=6.3~\gev$
respectively. The corresponding values in the HV scheme are
$\mu=7.5~\gev$ and $\mu=8.6~\gev$~.
\item
There is no value of $\mu=\mu_f$ in the full range considered for
which $\varepsilon_1(\mu)$ and $\varepsilon_8(\mu)$ or equivalently
$\xi_1^{\rm NF}(\mu)$ and $\xi_2^{\rm NF}(\mu)$ simultaneously
vanish. We also observe contrary to expectations in \cite{NS97}
that $\varepsilon_1(\mu)$ is not necessarily smaller than
$\varepsilon_8(\mu)$. In fact the large $N$ arguments presented
in \cite{NS97} that $\varepsilon_1(\mu)=\ord(1/N^2)$ and
$\varepsilon_8(\mu)=\ord(1/N)$, imply strictly speaking only
that the $\mu$-dependence of $\varepsilon_8(\mu)$
is much stronger than that of $\varepsilon_1(\mu)$,
which we indeed see in figs.~\ref{SILV1} and \ref{SILV2}. 
The hierarchy of their actual
values is a dynamical question. Even if the large $N$-counting rules
$\varepsilon_1(\mu)=\ord(1/N^2)$ and
$\varepsilon_8(\mu)=\ord(1/N)$ are true independently of the
factorization hypothesis \cite{EW,BGR}, it follows from our analysis
that once the generalized factorization hypothesis is made, the
extracted values of $\varepsilon_i$ violate for some range of $\mu$
the large-N rule $\varepsilon_1\ll\varepsilon_8$. 
\ei

\begin{table}[htb]
\caption[]{ $\xi^{\rm NF}_{1,2}(\mu)$ as functions of $\mu$
for different schemes and $\Lms^{(5)}=225\mev$.}
\label{tabf}
\begin{center}
\begin{tabular}{|c|c|c|c||c|c|c|}
\hline
& \multicolumn{3}{c||}{$\xi^{\rm NF}_1(\mu)$} &
  \multicolumn{3}{c| }{$\xi^{\rm NF}_2(\mu)$} \\
\hline
$\mu [{\rm GeV}]$ &NDR & HV & FS & NDR & HV & FS  \\
\hline
\hline
2.5 & 0.046 & 0.035 & --0.033 & 0.102 & 0.144 & 0.075 \\
\hline
5.0 & 0.065 & 0.059 & 0.001 & 0.022 & 0.055 &--0.004 \\
\hline
7.5 & 0.071 & 0.067 & 0.014 & --0.016 & 0.013 &--0.041 \\
\hline
10.0 & 0.074 & 0.071 & 0.021 &--0.039 & -0.013 &--0.064 \\
\hline
\end{tabular}
\end{center}
\end{table}

We can next investigate for which renormalization scheme characterized
by $\kappa_\pm$ the factorization is exact at $\mu_f=\mb=4.8~\gev$.
We call this choice the ``factorization scheme" (FS).
Using the central values in (\ref{aiexp}) and $\Lms^{(5)}=225\mev$
we find by means of
(\ref{kappa+}) and (\ref{kappa-})
\be\label{KPKM}
\kappa_+=13.5~, \qquad  \kappa_-=3.9~~~~~~~({\rm FS}).
\ee
These values deviate considerably from the NDR values $\kappa_\pm=0$
and the HV values $\kappa_\pm=\mp 4 $. Yet one can verify that for these
values $J_+=6.13 $ and $J_-=1.17$ and consequently in this scheme the NLO
corrections at $\mu=\mb$ remain perturbative.
In table \ref{tabf} we give the values of 
$\xi_i^{\rm NF}(\mu)$ for the NDR, HV and FS schemes. 

The numerical analysis presented here used as input the central values
for $a_i^{\rm eff}$ given in (\ref{aiexp}).
As stressed in particular in \cite{ITALY}, the strong model dependence
of the form factors and large experimental errors preclude at present
a precise determination of these parameters. Consequently when these
uncertainties are taken into account, the differences between
various schemes are washed out to some extent. Yet the general
features of the results obtained for other numerical values of the pair
$(a_1^{\rm eff},a_2^{\rm eff})$  are very similar to the ones
presented here. 

\section{Generalized Factorization and $N^{\rm eff}$}
As pointed sometime ago in \cite{BJLW1,rome2} and recently
discussed in \cite{Cheng,GNF,AKL98},
it is always possible
to calculate the  scale and scheme dependence of the hadronic matrix 
elements in perturbation theory by simply calculating the matrix elements
of the relevant operators between the quark states. 
Combining these scheme and scale dependent contributions with the
Wilson coefficients $C_i(\mu)$ one obtains the effective coefficients
$C_i^{\rm eff}$ which are free from these dependences. If one neglects
in addition final state interactions and other possible non-factorizable
contributions the decay amplitudes can be generally written as follows
\begin{equation}\label{ALI}
A=\langle H_{\rm eff}\rangle =\frac{G_F}{\sqrt{2}}V_{CKM}
\lbrack C_1^{\rm eff}\langle O_1\rangle^{\rm tree} +C_2^{\rm eff}
\langle O_2\rangle^{\rm tree}  \rbrack~,
\end{equation}
where $\langle O_i\rangle^{\rm tree}$ denote tree level matrix elements.
The proposal in \cite{Cheng,GNF,AKL98} is to use (\ref{ALI}) and to apply 
the idea of the factorization to the tree level matrix elements.
In this approach then the effective parameters $a_{1,2}^{\rm eff}$
are given by (\ref{BS23F}) with 
$N^{\rm eff}$  treated as a phenomenological parameter which
models those non-factorizable contributions to the hadronic matrix elements,
which have not been included into $C_i^{\rm eff}$.
In particular it has been suggested in \cite{Cheng,GNF,AKL98} that
the values for $N^{\rm eff}$ extracted from the data on two-body
non-leptonic decays should teach us about the pattern of 
non-factorizable contributions.

In particular when calculating the effective coefficients $C_i^{\rm eff}$, 
the authors of  \cite{GNF,AKL98}  have included a
subset of contributions to the perturbative matrix elements, which is
sufficient to cancel the scale and scheme dependence of the Wilson
coefficients.
Unfortunately the results of such calculations are generally gauge
dependent and suffer from the dependence on the infrared regulator
and generally on the assumptions about the external momenta. 

Let us discuss this point in detail, following \cite{rome2}.
The Green function of the renormalized operator $O$,
for a given choice of the ultraviolet regularization (NDR or HV for example), 
a choice of the external momenta $p$ and of the gauge parameter $\lambda$, 
is given by 
\begin{equation}
  \label{eq:matel}
  \Gamma_O^\lambda (p) = 1 + \frac{\alpha_s}{4 \pi} 
\left(-\frac{\gamma^{(0)}}{2}
  \ln(\frac{-p^2}{\mu^2}) + \hat{r} \right),
\end{equation}
with
\begin{equation}
\hat{r}= \hat r^{NDR,HV} + \lambda \hat r^\lambda.
\label{eq:rral}
\end{equation}
The matrices $\hat r^{NDR,HV}$ depend on the choice of the external 
momenta and
on the ultraviolet regularization, while $\hat r^\lambda$ is 
regularization- and
gauge-independent, but depends on the external momenta. 
It is clearly possible to define a renormalization scheme in which, 
for given external momenta and gauge parameter, $\Gamma_O^\lambda (p)
= 1$, or in other words $\langle O \rangle_{p,\lambda}=\langle O
\rangle^{\rm tree}$ (this corresponds to the RI scheme discussed in
\cite{rome2}). However, the definition of the renormalized 
operators will now depend on the choice of the gauge and of the
external momenta. If one were able, for example by means of lattice
QCD, to compute the matrix element of the operator using the same
renormalization prescription, the dependences on the gauge and on the
external momenta would cancel between the Wilson coefficient and the
matrix element. If, on the contrary, the matrix elements are estimated
using factorization, no trace is kept of the renormalization
prescription and the final result is gauge and infrared dependent. 

In  \cite{GNF,AKL98} scale- and scheme-independent effective
Wilson coefficients have been obtained by adding to $C_i(\mu)$ the
contributions coming from vertex-type quark matrix elements, denoted
by $\hat r_V$ and $\hat\gamma_V$. In particular
\begin{eqnarray}
  \label{eq:cali}
  C_1^{\rm eff}&=& 
C_1(\mu) + \frac{\alpha_s}{4 \pi}\left( r_V^T + \gamma_V^T
  \log \frac{m_b}{\mu}\right)_{1j} C_j(\mu),\nonumber \\
  C_2^{\rm eff}&=&
 C_2(\mu)  + \frac{\alpha_s}{4 \pi}\left( r_V^T + \gamma_V^T
  \log \frac{m_b}{\mu}\right)_{2j} C_j(\mu).
\end{eqnarray}
where the index $j$ runs through all contributing operators, also
penguin operators considered in \cite{Cheng,GNF,AKL98}.

 It is evident from the above discussion that $\hat r_V$  
depends not only on the
external momenta, but also on the gauge chosen.   
For example, in \cite{GNF,AKL98} the following result for $\hat r_V$ is
quoted:
\begin{equation}
  \label{eq:rvali}
  \hat r_V=\left(
    \begin{array}{cccccc}
      \frac{7}{3} & - 7 & 0 & 0 & 0 & 0\\
      - 7 & \frac{7}{3} & 0 & 0 & 0 & 0\\ 
      0 & 0 & \frac{7}{3} & - 7 & 0 & 0\\ 
      0 & 0 & - 7 & \frac{7}{3} & 0 & 0\\ 
      0 & 0 & 0 & 0 & - \frac{1}{3} & 1\\ 
      0 & 0 & 0 & 0 & - 3 & \frac{35}{3} 
    \end{array}
    \right).
\end{equation}
This result is valid in the Landau gauge ($\lambda=0$); 
in an arbitrary gauge, with the same choice of external momenta used
to obtain (\ref{eq:rvali}) one would get
\begin{equation}
  \label{eq:rvtot}
  \hat r_V = \hat r_V (\lambda=0) + \lambda r_V^\lambda,
\end{equation}
with $\hat r_V (\lambda=0)$ given in (\ref{eq:rvali}) and
\begin{equation}
  \label{eq:rvluca}
  r_V^\lambda=\left(
    \begin{array}{cccccc}
      - \frac{5}{6} & - \frac{3}{2} & 0 & 0 & 0 & 0\\
      - \frac{3}{2} & - \frac{5}{6} & 0 & 0 & 0 & 0\\
      0 & 0 & - \frac{5}{6} & - \frac{3}{2} & 0 & 0\\ 
      0 & 0 & - \frac{3}{2} & - \frac{5}{6} & 0 & 0\\ 
      0 & 0 & 0 & 0 & - \frac{11}{6} & \frac{3}{2}\\
      0 & 0 & 0 & 0 & 0 & \frac{8}{3}
    \end{array}
    \right).
\end{equation}
The expressions for the full 
$10 \times 10$ $\hat r$ matrices in the
NDR and HV schemes and in the Feynman and Landau gauges are given in
\cite{rome2}, for a different choice of the external momenta.
The results for the Landau gauge are given in \cite{BJLW1}.

Equation (\ref{eq:rvtot}) shows that the definition of the effective
coefficients advocated in \cite{Cheng,GNF,AKL98}
is gauge-dependent. In addition, it also depends on the choice of the
external momenta.
This implies that the effective number of colors extracted in
\cite{Cheng,GNF,AKL98} is also gauge-dependent, and therefore it
cannot have any physical meaning. 

The gauge dependences and infrared dependences discussed here are not
new. They appear in any calculation of matrix elements of operators
between quark states necessary in the process of matching of the
full theory onto an effective theory. A particular example can be
found in \cite{BJW90} where the full gauge dependence of the quark
matrix element of the operator $(\bar s d)_{V-A}(\bar s d)_{V-A}$
has been calculated. However, in the process of matching such
unphysical dependences in the effective theory are cancelled by
the corresponding contributions in the full theory so that the
Wilson coefficients are free of such dependences. Similarly
in the case of inclusive decays of heavy quarks, where the spectator
model can be used, they are cancelled by gluon
bremsstrahlung. In exclusive hadron decays there is no meaningful way to
include such effects in a perturbative framework and one is left
with the gauge and infrared dependences in question.

\section{Summary}
In this paper we have critically analyzed the hypothesis of
the generalized factorization. While the parametrization of the data
in terms of a set of effective parameters discussed in
\cite{Cheng}--\cite{AKL98}
 may appear to be useful,
we do not think that this approach offers convincing means to
analyze the physics of non-factorizable contributions to
non-leptonic decays. In particular:
\begin{itemize}
\item
The renormalization scheme dependence of the non-factorizable
contributions to hadronic matrix elements precludes the
determination of the factorization scale $\mu_f$.
\item
Consequently for any chosen value of $\mu_f=\ord(m_b)$ 
it is possible to find a renormalization
scheme for which the non-perturbative parameters $\varepsilon_{1,8}$
used in \cite{NS97} to characterize the size of non-factorizable
contributions vanish. The same applies to 
$\xi^{\rm NF}_{1,2}(\mu)$ introduced in the present paper.
\item
We point out that the recent extractions of the effective number
of colours $N^{\rm eff}$ from two-body non-leptonic B-decays,
presented in \cite{Cheng,GNF,AKL98},
while $\mu$ and renormalization scheme independent suffer from
gauge dependences and infrared regulator dependences.
\end{itemize}

A further problem in the generalized factorization approach
is given by the presence in many channels of operators that
contribute only through non-factorizable terms. These contributions
cannot be incorporated in the definitions of $\R1$ and $\E8$, and
a more general parametrization is needed \cite{CFMS1}.
A typical example is given by charming-penguin contributions to
$B\to K\pi$ decays \cite{CFMS2}.

We hope that our analysis demonstrates clearly the need for
an approach to non-leptonic decays which goes
beyond the generalized factorization discussed recently in
the literature. Some possibilities are offered by dynamical approaches
like QCD sum rules as recently reviewed in \cite{KR98}. 
However, even a phenomenological approach which does not suffer
from the weak points of factorization discussed here, would
be a step forward. We hope to present some ideas in this direction
in a forthcoming publication \cite{BUSI2}.

\vfill\eject
 

\begin{thebibliography}{99}
\bibitem{FEYNMAN}
J. Schwinger, { Phys. Rev. Lett.} {\bf 12} (1964) 630; 
R.P. Feynman, in {\it Symmetries in Particle Physics}, ed. A. Zichichi,
Acad. Press 1965, p.167; O. Haan and B. Stech, 
{ Nucl. Phys.} {\bf B 22}  (1970) 448.  
\bibitem{STECHF}
D. Fakirov and B. Stech, { Nucl. Phys.} {\bf B 133}  (1978) 315;
L.L. Chau, Phys. Rep. {\bf 95} (1983) 1.  
\bibitem{BAUER}
M. Wirbel, B. Stech and M. Bauer, { Z. Phys.} {\bf C 29} (1985) 637;
M. Bauer, B. Stech and M. Wirbel, { Z. Phys.} {\bf C 34} (1987) 103.
\bibitem{NEUBERT}
M. Neubert, V. Rieckert, B. Stech and Q.P. Xu, in ``Heavy Flavours",
 eds. A.J. Buras and M. Lindner (World Scientific, Singapore, 1992),
p. 286;
\bibitem{Cheng}
H.-Y. Cheng, {Phys. Lett.} {\bf B 335} (1994) 428,
{ Phys. Lett.} {\bf B 395} (1997) 345;
H.-Y. Cheng and B. Tseng, hep-ph/9708211, hep-ph/9803457.
\bibitem{Soares}
J.M. Soares, { Phys. Rev.} {\bf D 51} (1995) 3518.
\bibitem{NS97}
M. Neubert and B. Stech,
Preprint CERN-TH/97-99, hep-ph/9705292, to appear in Heavy Flavours II,
edited by A.J. Buras and M. Lindner (World Scientific, Singapore);
B. Stech, hep-ph/9706384;
M.Neubert, { Nucl. Phys. Proc. Suppl.} {\bf B 64} (1998) 474, 
hep-ph/9801269.
\bibitem{GNF}
A. Ali and C. Greub, { Phys. Rev.} {\bf D57} (1998) 2996;
A. Ali, J. Chay, C. Greub and P. Ko,
 { Phys. Lett.} {\bf B 424} (1998) 161.
\bibitem{AKL98}
A. Ali, G. Kramer and C.-D. L\"u, hep-ph/9804363.
\bibitem{LNF}
D. Du and Z. Xing, { Phys. Lett.} {\bf B 312} (1993) 199;
A. Deandrea et al., { Phys. Lett.} {\bf B 318} (1993) 549,
{ Phys. Lett.} {\bf B 320} (1994) 170;
N.G. Deshpande, B. Dutta, S. Oh, { Phys. Rev.} {\bf D 57} (1998) 5723, 
hep-ph/9712445.
\bibitem{AJBNF}
A.J. Buras, 
{ Nucl. Phys.} {\bf B 434} (1995) 606.
\bibitem{BJL}
A.J. Buras, M. Jamin, and M.E. Lautenbacher, 
{ Nucl. Phys.} {\bf B 408} (1993) 209.
\bibitem{BJLW1}
{ A.J. Buras, M. Jamin, M.E. Lautenbacher and P.H. Weisz,}
{ Nucl. Phys.} {\bf B 370} (1992) 69,
{ Nucl. Phys.} {\bf B 400} (1993) 37.
\bibitem{BJLW2}
{ A.J. Buras, M. Jamin and M.E. Lautenbacher,}
{ Nucl. Phys.} {\bf B 400} (1993) 75.
\bibitem{ROME}
{ M. Ciuchini, E. Franco, G. Martinelli and L. Reina,}
{ Phys. Lett.} {\bf B 301} (1993) 263,
{ Nucl. Phys.} {\bf B 415} (1994) 403.
\bibitem{WEISZ}
A.J. Buras and P.H. Weisz,
{ Nucl. Phys.} {\bf B 333} (1990) 66.
\bibitem{ALT}
 G. Altarelli, G. Curci, G. Martinelli and S. Petrarca,
{ Nucl. Phys.} {\bf B 187} (1981) 461.
\bibitem{BBDM}
W.A. Bardeen, A.J. Buras, D.W. Duke and T. Muta,
{ Phys. Rev.} {\bf D 18} (1978) 3998.
\bibitem{WEBER}
M. Schmelling, in proceedings of the 28th International Conference on
High Energy Physics, July 1996, Warsaw, Poland, page 91.
\bibitem{EW}
 E. Witten,
{ Nucl. Phys.} {\bf B 160} (1979) 57.
\bibitem{BGR}
 A.J. Buras, J.M. G\`erard and R. R\"uckl,
{ Nucl. Phys.} {\bf B 268} (1986) 16.
\bibitem{ITALY}
M. Ciuchini, R. Contino, E. Franco, G. Martinelli and L. Silvestrini,
hep-ph/9801420.
\bibitem{rome2}
M. Ciuchini, E. Franco, G. Martinelli, L. Reina and L. Silvestrini, 
Z.Phys. {\bf C68} (1995) 239.
\bibitem{BJW90}
A.J. Buras, M. Jamin, and P.H. Weisz,
{ Nucl. Phys.} {\bf B 347} (1990) 491.
\bibitem{CFMS1}
M. Ciuchini, E. Franco, G. Martinelli and L. Silvestrini, 
{ Nucl. Phys.} {\bf B 501} (1997) 271.
\bibitem{CFMS2}
M. Ciuchini, R. Contino, E. Franco, G. Martinelli and L. Silvestrini, 
{ Nucl. Phys.} {\bf B 512} (1998) 3.
\bibitem{KR98}
A. Khodjamirian and R. R\"uckl, hep-ph/9801443,
to appear in Heavy Flavours II,
edited by A.J. Buras and M. Lindner (World Scientific, Singapore).
\bibitem{BUSI2}
A.J. Buras and L. Silvestrini, work in progress.
\end{thebibliography}
\end{document}